# Gravitational Radiation from String Cosmology

R. Brustein[a,c,*], M. Gasperini[b], M. Giovannini[b,c] and G. Veneziano[c]

[a] *Department of Physics, Ben-Gurion University, Beer-Sheva 84105, Israel*
[b] *Dipartimento di Fisica Teorica, Via P.Giuria 1, 10125 Turin, Italy*
[c] *Theory Division, CERN, CH-1211, Geneva 23, Switzerland*

## Abstract

A spectrum of relic stochastic gravitational radiation, strongly tilted towards high frequencies, and characterized by two basic parameters is shown to emerge in a class of string theory models. We estimate the required sensitivity for detection of the predicted gravitational radiation and show that a region of our parameter space is within reach for some of the planned gravitational-wave detectors.

Cosmological predictions of string theory originate from physics of the Early-Universe, when space-time curvatures may have been of Planckian strength and therefore have a chance to produce observable effects. In this talk we present such an effect in the form of produced gravitational radiation. The results presented in this talk were obtained in [1] where more details and references can be found.

We shall consider a class of models in which a period of dilaton-driven inflation [2,3] is followed by a stringy epoch, during which the curvature remains of the order of the string scale and later by the standard (radiation then matter dominated) cosmology. As discussed in detail elsewhere [4,5], the presence of a high-curvature stringy epoch appears to be unavoidable for a viable inflationary string cosmology scenario. We [6] showed that both scalar and tensor perturbations exhibit very similar spectra, which, unlike the spectra of the standard inflationary scenarios, are not flat but strongly tilted towards higher frequencies, as originally noted [7].

Let us consider an isotropic, $(3+1)$-dimensional, spatially flat cosmology. Following [7] (see also [6]) it is easy to show that the Fourier modes $\psi_k$ of each of the two canonical variables associated with physical, transverse-traceless, polarizations of tensor perturbations satisfy, in the string frame, the following simple wave equation

$$\psi_k'' + [k^2 - V(\eta)]\psi_k = 0, \qquad V(\eta) = (g/a)(a/g)'' \tag{1}$$

where a prime denotes differentiation with respect to conformal time $\eta$ ($ad\eta \equiv dt$), $k$ is the comoving wave number related to the physical one, $\omega$, by $k = \omega a$, $a(t)$ is the isotropic scale factor, and $g = \exp(\phi/2)$.

A given mode $k$ will be well inside the horizon initially, then hit the potential barrier $V(\eta)$ at some "exit" time $\eta_{ex} \sim k^{-1}$, and leave the barrier at some later "reentry" time $\eta = \eta_{re} \sim \eta_1$. The approximate solutions of eq. (1), normalized to vacuum fluctuations, in these three regimes are given by

$$\psi_k = \frac{\lambda_s}{\sqrt{k}} e^{-ik\eta}, \qquad \eta < \eta_{ex} \tag{2}$$

$$\psi_k = \frac{a}{g}\left[A_k + B_k \int^\eta d\eta' \left(\frac{g}{a}\right)^2\right], \qquad \eta_{ex} < \eta < \eta_{re} \tag{3}$$

$$\psi_k = \frac{\lambda_s}{\sqrt{k}}\left[c_+(k)e^{-ik\eta} + c_-(k)e^{ik\eta}\right], \qquad \eta > \eta_{re} \tag{4}$$

In (4), the magnitude of $c_-$ gives the amplification of the GW with respect to a minimal vacuum fluctuation and it can be obtained by matching the above solutions and their first derivatives at each transition time,

$$\begin{aligned} 2k\eta_1|c_-(k)| &\simeq \frac{g_{ex}/a_{ex}}{g_{re}/a_{re}}\left[1 + k\eta_1\left(\frac{g_{re}/a_{re}}{g_{ex}/a_{ex}}\right)^2 \right. \\ &\left. + k(g_{ex}/a_{ex})^{-2}\int_{\eta_{ex}}^{\eta_1} d\eta\, (g/a)^2\right]. \end{aligned} \tag{5}$$

The dilaton-driven inflationary background is given by [3] $a(\eta) = (-\eta)^{-\frac{1}{1+\sqrt{3}}}$, $\phi(\eta) = -\sqrt{3}\ln(-\eta)$, $a/g \sim (-\eta)^{1/2}$, $-\infty < \eta < 0$ while, for the string era, we will assume that $H$ and $\partial_t\phi$ are approximately constant. For those scales which crossed the horizon during the dilaton-driven phase, we arrive at the following estimate for the r.m.s. perturbation amplitude over a comoving length scale $k^{-1}$ which is given, in general, by $|\delta h_k(\eta)| \simeq k^{3/2}|h_k| = (g/a)k^{3/2}|\psi_k|$. For $\eta > \eta_1$, we find in terms of present, red-shifted proper frequencies $\omega = k/a$

$$\begin{aligned} |\delta h_\omega| &\simeq \sqrt{\frac{H_0}{M_s}} z_{eq}^{-1/4} g_s z_s \left(\frac{\omega}{\omega_s}\right)^{1/2} \\ &\quad \left[1 + \frac{1}{2}\ln\left(\frac{\omega_s}{\omega}\right) + z_s^{-3}\left(\frac{g_1}{g_s}\right)^2\right], \quad \omega < \omega_s \end{aligned} \tag{6}$$

$$\begin{aligned} \omega_s &= k_s/a \simeq z_{eq}^{-1/4}\sqrt{H_0 M_s} z_s^{-1} \\ &\equiv z_s^{-1}\omega_1 \sim z_s^{-1} g_1^{1/2} 10^{11} Hz \end{aligned} \tag{7}$$

where $z_{eq} = a/a_{eq} \sim 10^4$ takes into account the transition from radiation to matter dominance at $t = t_{eq}$, $\omega_1 = H_1 a_1/a \sim 10^{11}$Hz is the maximal frequency reached during the string phase, $M_s = M_{Pl} g = \lambda_s^{-1} \sim H_1$, and $H_0 \sim 10^{-18}$Hz is the present value of the Hubble scale.

The fraction of critical density, $\Omega_{GW} = \rho_{GW}/\rho_c$, stored in our GW per logarithmic interval of $\omega$, defined by $d\Omega_{GW}/(d\ln\omega) = \omega^4 |c_-|^2/(M_p H)^2$ is given by

$$\frac{d\Omega_{GW}}{d\ln\omega} = z_{eq}^{-1} g_s^2 \left(\frac{\omega}{\omega_s}\right)^3 \left[1 + \frac{1}{2}\ln\left(\frac{\omega_s}{\omega}\right) + z_s^{-3}\left(\frac{g_1}{g_s}\right)^2\right]^2 \quad (8)$$

$$\sim \left(\frac{\omega}{H_0}\right)^2 |\delta h_\omega|^2, \quad \omega < \omega_s$$

Finally we give, with the appropriate caveats, the generalization of the above results to frequencies whose exit occurred during the stringy phase,

$$|\delta h_\omega| \simeq g_1 \sqrt{\frac{H_0}{M_s}} z_{eq}^{-1/4} \left[\left(\frac{\omega}{\omega_1}\right)^{2-\beta} + \left(\frac{\omega}{\omega_1}\right)^{\beta-1}\right] \quad (9)$$

$$\frac{d\Omega_{GW}}{d\ln\omega} \simeq g_1^2 z_{eq}^{-1} \left[\left(\frac{\omega}{\omega_1}\right)^{6-2\beta} + \left(\frac{\omega}{\omega_1}\right)^{2\beta}\right], \quad \omega_s < \omega < \omega_1 \quad (10)$$

where $\beta = -\log(g_s/g_1)/\log z_s$ is the average value of $\dot{g}/(gH)$.

We would like to discuss now the prospects of observing our spectrum in gravitational wave detectors. Our main emphasis will be on the planned large interferometers LIGO [8] and VIRGO [9], which are expected [10] to start operating at sensitivities of $d\Omega_{GW}/d\ln\omega = 10^{-6}$ in a frequency band around a few hundred Hz, and have set the ambitious final sensitivity goals of $d\Omega_{GW}/d\ln\omega = 10^{-10}$ in a frequency band around $\omega_I = 100 Hz$.

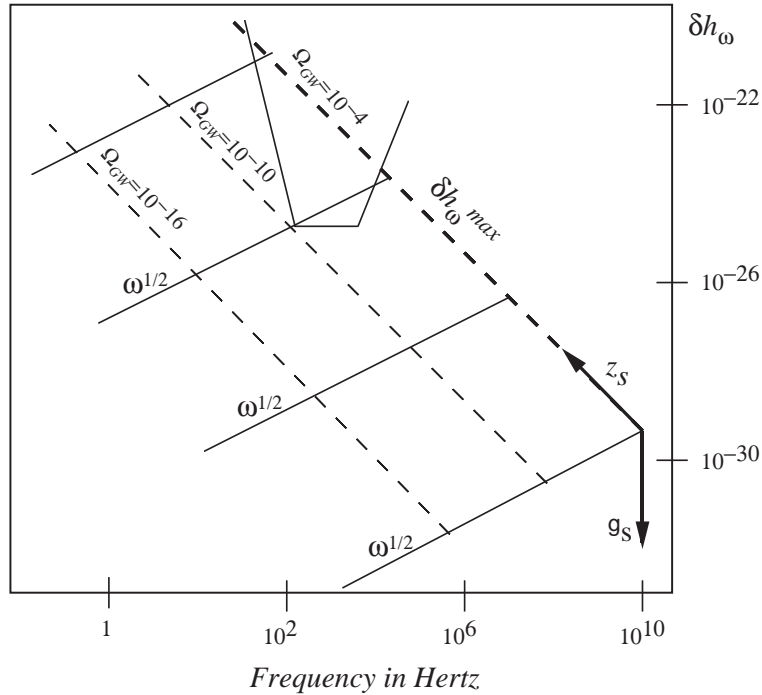

The spectral amplitude of gravitational waves $|\delta h_\omega|$. The solid lines show several individual spectra for different values of $z_s$ and $g_s = 1$. The thick dashed line shows the maximum amplitude as a function of $z_s$ for $g_s = 1$. The dashed lines are lines of fixed $g_s$ and therefore lines of constant energy density. $\Omega_{GW}$ is roughly the maximal amount of gravitational energy density at a given value of $g_s$. Also shown are the sensitivity goals for the detection of stochastic background of the "Advanced LIGO".

From eq. (7) we can immediately see that the accessible region requires large values of $z_s$. We have to impose the bound following from pulsar-timing measurements [11], which implies $d\Omega_{GW}/d\ln\omega \lesssim 10^{-6}$ at $\omega_P = 10^{-8} Hz$. We also accept the bound $\Omega_{GW} \lesssim 0.1$, imposed by standard nucleosynthesis [12]. Moreover, for consistency, we impose that the amplified perturbations have a negligible back-reaction on the metric, namely $d\Omega_{GW}/d\ln\omega < 1$ at all frequencies and times. In **Figure 1** the predicted spectrum of gravitational waves as a function of frequency, taking into account the above constraints, is depicted for $\beta < 3/2$, $z_s < 10^9$, in terms of the quantity $|\delta h_\omega|$ (denoted $h_c$ in [10] ), which represents the characteristic amplitude of a stochastic background. Also shown, for comparison the sensitivity goals of the "Advanced Ligo" in terms of the quantity $h_{3/yr}$ defined as the amplitude necessary for detection of a stochastic background at the 90% confidence level in a 1/3 of a year (see [10] for exact definitions).

It may well be that new detectors based on old ideas[13], or even other detectors, will be more suitable for detecting our spectra.